\documentclass[conference]{IEEEtran}
\IEEEoverridecommandlockouts
\usepackage{cite}
\usepackage{amsmath,amssymb,amsfonts}
\usepackage{algorithmic}
\usepackage{graphicx}
\usepackage{textcomp}
\usepackage{xcolor}
\usepackage{orcidlink}
\usepackage[caption=false,font=footnotesize]{subfig}
\usepackage{url}
\usepackage{array}

\def\BibTeX{{\rm B\kern-.05em{\sc i\kern-.025em b}\kern-.08em
    T\kern-.1667em\lower.7ex\hbox{E}\kern-.125emX}}
\begin{document}

\title{Curiosity Over Hype: Modeling Motivation Language to Understand Early Outcomes in a Selective Quantum Track}

\author{
\IEEEauthorblockN{Daniella Alexandra Crysti Vargas Saldaña}
\IEEEauthorblockA{\textit{School of Engineering} \\
\textit{Universidad Peruana de Ciencias Aplicadas}\\
Lima, Peru \\
ORCID: 0009-0001-6926-2469}
\and
\IEEEauthorblockN{Freddy Herrera Cueva}
\IEEEauthorblockA{\textit{Center for Quantum Networks} \\
\textit{NSF Engineering Research Center}\\
Maryland, USA \\
ORCID: 0009-0009-4793-7195}
}

\IEEEpubid{\makebox[\columnwidth]{979-8-3315-8153-4/25/\$31.00~\copyright~2025 IEEE\hfill}
\hspace{\columnsep}\makebox[\columnwidth]{}}
\maketitle

\maketitle

\begin{abstract}
We study whether latent motivation signals in short Spanish admission responses predict engagement and performance in an early quantum computing pathway run by \emph{QuantumHub Perú}. We analyze \(N=241\) applicants’ open responses and link them to outcomes from two selective modules: Module~1 (secondary; mathematics and computing foundations; \(n=23\)) and Module~2 (secondary + early undergraduate; quantum fundamentals; \(n=36\), including M1 continuers). To ensure baseline comparability, the M2 university entrance exam matched the difficulty of the M1 final.

Final grades followed the program’s official cohort-specific weightings (attendance/assignments/exam), which we retain to preserve ecological validity. Methodologically, we model text with Latent Dirichlet Allocation (LDA, \(k=8\)) \emph{and}, for robustness, with sentence embeddings from a \emph{small multilingual language model}, EmbeddingGemma-300M, projected via UMAP and clustered with HDBSCAN. This combination leverages the transparency of bag-of-words topics and the semantic richness of small language model (SLM) embeddings.

Descriptively, curiosity/learning topics show higher grades and attendance than technology/career-oriented topics; inferential tests are underpowered (e.g., linear \(R^2 \approx 0.03\); logistic pseudo-\(R^2 \approx 0.04\)) so effect-size estimates should be viewed as preliminary rather than confirmatory. Embedding-based clustering yields seven clusters with 11.2\% noise and modest agreement with LDA (ARI \(=0.068\); NMI \(=0.163\)). Results suggest that brief motivation responses encode promising signals that could support early mentoring in rigorous STEM pipelines, while highlighting the need for larger, pre-registered studies.
\end{abstract}


\section{Introduction}
\IEEEpubidadjcol

Motivation is central to persistence and achievement in rigorous STEM pathways. Yet scalable indicators often require long surveys or interviews. A ubiquitous but underused signal in selective programs is the short admission response where applicants state why they wish to enroll. In emerging domains like quantum computing education---with limited evidence in school-age contexts, especially in Latin America~\cite{gragera2025global,maldonadoromo2023quantum}---low-cost text signals are attractive.

We analyze Spanish short responses from applicants to \emph{QuantumHub Perú}, which runs a two-step pipeline: \textbf{Module 1 (M1)} for secondary students (mathematics/computing foundations) and \textbf{Module 2 (M2)} for secondary and early undergraduates (introductory quantum theory and computation). The M2 university exam matched the difficulty of the M1 final to align academic thresholds. We operationalize motivation using unsupervised topics and dense embeddings, connect these signals to grades/attendance, and interpret themes along an \emph{intrinsic} (curiosity/learning) vs.\ \emph{instrumental} (technology/career) axis~\cite{ryan2000sdt,eccles2002expectancy}.

Our contribution is threefold. First, we provide early evidence that short Spanish motivation statements in a selective quantum pathway can be organized into latent themes aligned with classic motivational theories. Second, we combine transparent topic modeling (LDA) with small language model (SLM) embeddings from a modern multilingual encoder, EmbeddingGemma-300M, illustrating a portable pipeline suitable for resource-constrained educational settings. Third, we link these signals to early outcomes, showing descriptive but non-significant trends that motivate larger, pre-registered studies.

\section{Methods}
\subsection{Setting}
\emph{QuantumHub Perú} runs a selective early pipeline in quantum computing with two modules: \textbf{M1} (secondary; mathematics \& computing foundations) and \textbf{M2} (secondary + early undergraduate; quantum fundamentals). To align baselines across cohorts, the M2 university entrance exam matched the difficulty of the M1 final.

\subsection{Participants and Data}
We analyzed \(N=241\) applicants who submitted short open responses in Spanish at admission. Academic outcomes were available for \(n=23\) in M1 and \(n=36\) in M2 (including M1 continuers). Text was lower-cased, accent-stripped, tokenized, and stop-word filtered (Spanish).

\subsection{Text Representations: Bag-of-Words vs.\ Small LM Embeddings}
\paragraph*{Motivation for a hybrid approach.}
Bag-of-words models (e.g., LDA) are transparent and useful for discovering themes, but they ignore word order and morphology and struggle with synonymy, dialectal variation (ES-PE vs.\ ES-LA), and short answers. To complement LDA, we therefore used sentence embeddings from a \emph{small language model (SLM)} that maps semantically similar texts into nearby vectors even when vocabulary differs. Modern SLM encoders inherit many strengths of large language models while remaining lightweight enough for “at-home” educational research.

\paragraph*{(a) Topics (LDA).}
We fit Latent Dirichlet Allocation~\cite{blei2003lda} with \(k=8\) topics on a count matrix built with unigrams/bigrams, \(\texttt{min\_df}=5\), \(\texttt{max\_df}=0.8\). For each applicant we stored (i) the topic-proportion vector \(\boldsymbol{\theta}\) and (ii) the dominant-topic label.

\paragraph*{(b) Small multilingual LM embeddings.}
We computed sentence embeddings with \emph{google/embeddinggemma-300m} (released on the Hugging Face Hub), a compact multilingual encoder that produces dense vectors aligned across languages. We projected the embeddings using UMAP (cosine metric; \(n_{\text{neighbors}}=5\), \(n_{\text{components}}=30\)) and clustered with HDBSCAN (EOM; \(\text{min\_cluster\_size}=8\), \(\text{min\_samples}=1\)). We summarize internal validity with the silhouette score and compare agreement with LDA using ARI/NMI. This SLM layer serves as a robustness check: if clusters cohere semantically and partially align with LDA, it strengthens confidence that results are not artifacts of tokenization.

\subsection{Outcomes}
Final grades follow the official cohort-specific formulas summarized in
Table~\ref{tab:modules}. Attendance is a percentage; grades are on a 0–100 scale.
We also define \textit{passed\_any} (passed M1 and/or M2).
\begin{table}[t]
  \centering
  \caption{Overview of modules and grading formulas.}
  \label{tab:modules}
  \begin{tabular}{llp{0.22\columnwidth}p{0.27\columnwidth}}
    \hline
    \textbf{Module} & \textbf{Level} & \textbf{Content focus} & \textbf{Final grade} \\
    \hline
    M1 & Secondary &
    Math \& computing foundations &
    50\% attendance, 25\% assignments, 25\% exam \\
    M2 (univ.) & Early undergrad &
    Quantum fundamentals &
    25\% attendance, 25\% assignments, 50\% exam \\
    M2 (sec.) & Secondary &
    Quantum fundamentals &
    40\% attendance, 15\% assignments, 45\% exam \\
    \hline
  \end{tabular}
\end{table}

\subsection{Optional Covariate: BFI-10 Openness}
A subset of students completed a brief Big Five inventory (BFI-10). We compute \emph{Openness} as the mean of two items, reversing ``has few artistic interests'' and keeping ``has an active imagination'' as is. Because this covariate is available only for a small sample, we report it descriptively and include it as an optional control in sensitivity analyses.

\subsection{Analyses}
We conducted the following analyses:
\begin{enumerate}
  \item topic frequencies;
  \item descriptive means for grade and attendance by dominant topic;
  \item linear regression for M2 grade \(\sim\) topic proportions + cohort;
  \item logistic regression for \textit{passed\_any};
  \item ANOVA/Kruskal–Wallis tests across dominant topics.
\end{enumerate}
The SLM-based clustering (silhouette, ARI/NMI vs.\ LDA) is used as a
robustness check for semantic structure.

\section{Results}
\subsection{Topic Structure}
LDA with \(k=8\) yielded a compact, interpretable structure that maps naturally
onto an intrinsic–instrumental axis. Table~\ref{tab:topics} summarizes each
topic with a brief label and its dominant orientation. 

\begin{table}[t]
  \centering
  \caption{LDA topics: labels and dominant orientation.}
  \label{tab:topics}
  \begin{tabular}{cll}
    \hline
    \textbf{Topic} & \textbf{Short label} & \textbf{Orientation} \\
    \hline
    T0 & Problem-solving / physics / complexity & Mixed \\
    T1 & Learning / knowledge growth            & Intrinsic \\
    T2 & Curiosity / ``I'd like to learn''      & Intrinsic \\
    T3 & Specialization / systems               & Mixed \\
    T4 & Technology / future / ``I want to''    & Instrumental \\
    T5 & Course / career orientation            & Instrumental \\
    T6 & Opportunity / new field                & Instrumental \\
    T7 & Understanding ``how it works''         & Intrinsic \\
    \hline
  \end{tabular}
\end{table}

\begin{figure}[t]
  \centering
  \includegraphics[width=\columnwidth]{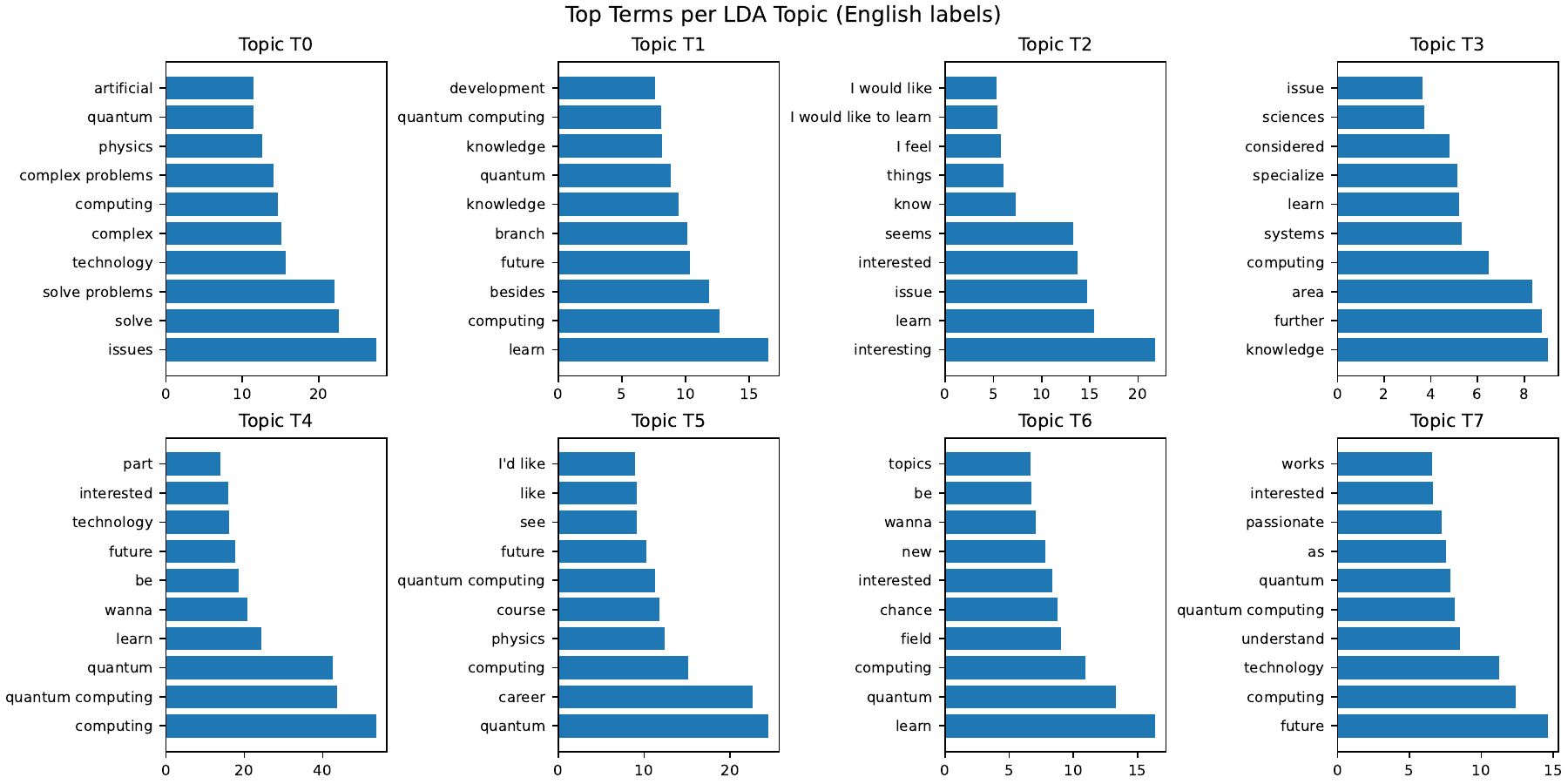}
  \caption{Top terms per LDA topic (T0--T7), translated to English.}
  \label{fig:topic-bars}
\end{figure}

Figure~\ref{fig:topic-bars} shows the highest-weight terms per topic
(translated to English), providing face validity for the intended semantics.

As shown in Table~\ref{tab:topics}, topics T1--T2 and T7 capture
curiosity- and understanding-oriented language, while T4--T6 reflect
more technology- and career-oriented framings.

\paragraph*{Prevalence.}
Table~\ref{tab:prevalence} reports dominant-topic counts and proportions
across the \(N=241\) applicants.

\begin{table}[t]
  \centering
  \caption{Dominant-topic prevalence across applicants.}
  \label{tab:prevalence}
  \begin{tabular}{lrr}
    \hline
    \textbf{Topic} & \textbf{Count} & \textbf{Share (\%)} \\
    \hline
    T4 & 44 & 18.3 \\
    T0 & 39 & 16.2 \\
    T5 & 36 & 14.9 \\
    T2 & 31 & 12.9 \\
    T1 & 25 & 10.4 \\
    T6 & 24 & 10.0 \\
    T7 & 24 & 10.0 \\
    T3 & 18 & 7.5  \\
    \hline
  \end{tabular}
\end{table}

Intrinsically oriented topics (T1--T2) are frequent but not dominant, while
more instrumental framings (T4--T5) jointly account for nearly one third of
responses, consistent with a selective STEM applicant pool where curiosity,
future orientation, and perceived opportunity co-exist.

\subsection{Performance by Topic (Descriptives)}
\begin{figure}[t]
  \centering
  \includegraphics[width=\columnwidth]{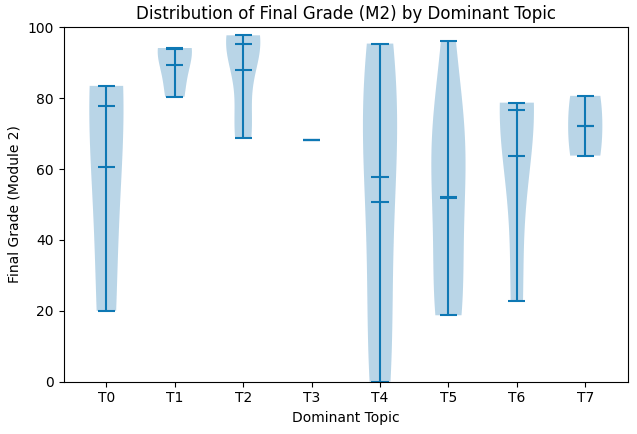}
  \caption{Distribution of M2 final grade by dominant topic (0--100).}
  \label{fig:m2-violin}
\end{figure}

Linking topics to outcomes, mean M2 grades were highest for curiosity/learning topics \(\mathbf{T1}\) and \(\mathbf{T2}\) (\(\approx 89.4\%\) and \(87.9\%\)), moderate for understanding/``how it works'' \(\mathbf{T7}\) (\(\approx 72.2\%\)), and lowest for technology/career-oriented \(\mathbf{T4}\)–\(\mathbf{T5}\) (\(\approx 50.7\%\)–\(51.7\%\)). Figure~\ref{fig:m2-violin} shows the full distribution of M2 grades by dominant topic: dispersion is wider for T0 and especially T4–T5 (longer lower tails), whereas T1–T2 concentrate in the upper range with shorter tails. M1 means are reported for completeness (\(n=23\)) and show ceiling effects in attendance (e.g., M1 T5 \(\approx 100\%\)), which cautions interpretation of attendance differences.

\subsection{Inferential Tests}
We next modeled M2 performance using continuous topic proportions (thus using all topical information per essay) and a cohort control. \textbf{Linear regression} (M2 grade \(\sim\) topic proportions + cohort) yields \(R^2 = 0.029\) (small). The coefficient for \(\mathbf{T2}\) (curiosity) is positive, while \(\mathbf{T6}\)–\(\mathbf{T7}\) trend negative; the cohort indicator (university track) is small and non-significant. \textbf{Logistic regression} for \emph{passed\_any} shows pseudo-\(R^2 \approx 0.038\); signs mirror the linear model but coefficients are non-significant. \textbf{Group comparisons} across dominant topics are non-significant (ANOVA \(F(7,\cdot)=1.58,\,p=.182\); Kruskal–Wallis \(H=11.29,\,p=.127\)). Given the limited outcome samples and heterogeneous grading formulas by cohort, these non-rejections are consistent with low power rather than evidence of no association.

\begin{figure}[t]
  \centering
  \includegraphics[width=\columnwidth]{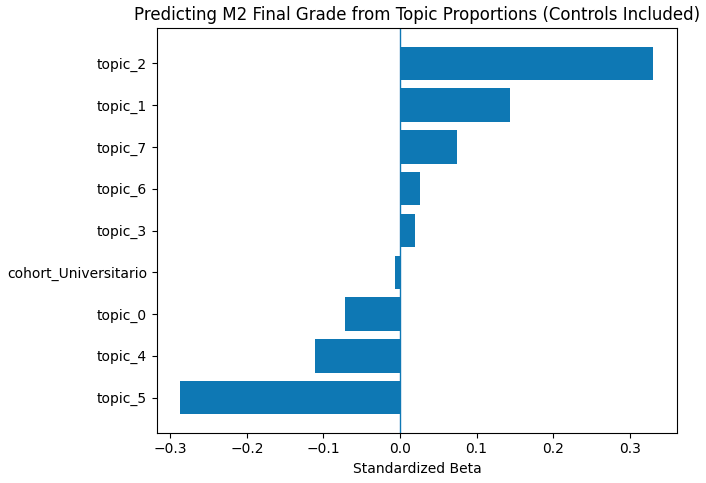}
  \caption{Standardized coefficients for predicting M2 grade from topic proportions (cohort control included).}
  \label{fig:coefplot}
\end{figure}

\subsection{Power Considerations}
A post-hoc power approximation indicates that detecting even medium effects (\(f \approx 0.25\)) would require outcome samples on the order of \(n \approx 120\)–\(150\) per cohort, far above our available outcome sample sizes (M1 \(n=23\); M2 \(n=36\)). Therefore, non-significant results are consistent with insufficient power, not with evidence of null relationships.

\subsection{Embedding-based Clustering (Robustness)}
\begin{figure*}[t]
  \centering
  \subfloat[T0]{\includegraphics[width=0.22\textwidth]{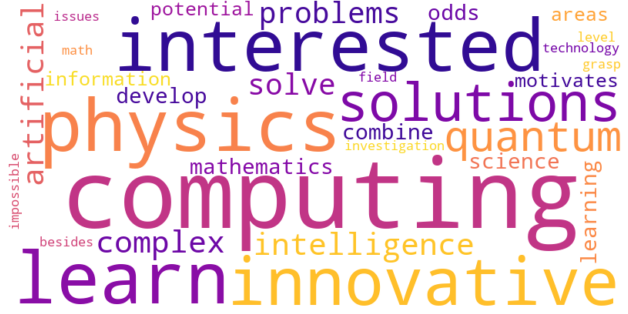}}\hfill
  \subfloat[T1]{\includegraphics[width=0.22\textwidth]{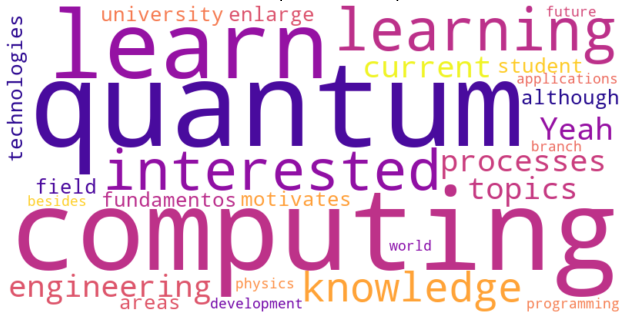}}\hfill
  \subfloat[T2]{\includegraphics[width=0.22\textwidth]{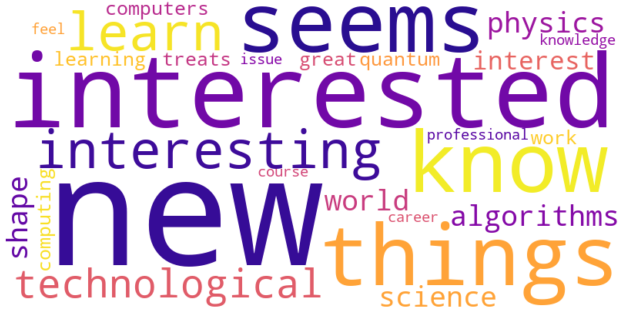}}\hfill
  \subfloat[T3]{\includegraphics[width=0.22\textwidth]{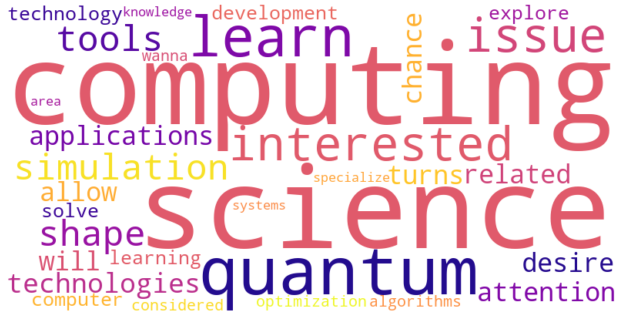}}\\[4pt]
  \subfloat[T4]{\includegraphics[width=0.22\textwidth]{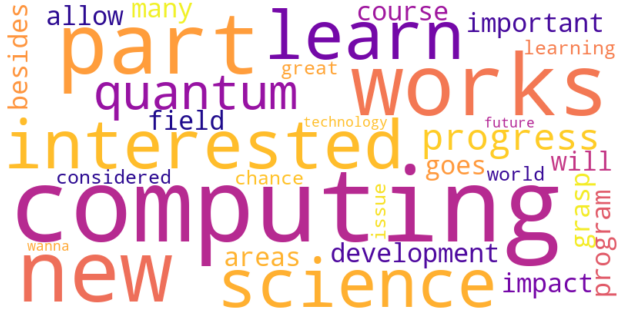}}\hfill
  \subfloat[T5]{\includegraphics[width=0.22\textwidth]{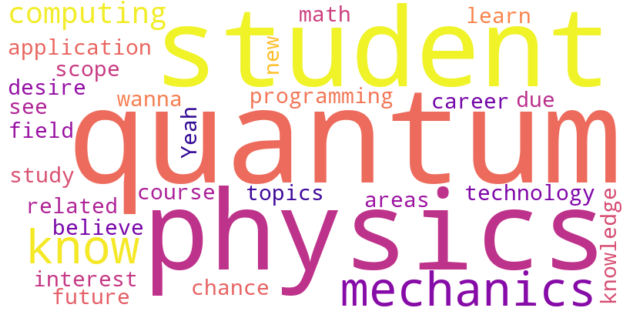}}\hfill
  \subfloat[T6]{\includegraphics[width=0.22\textwidth]{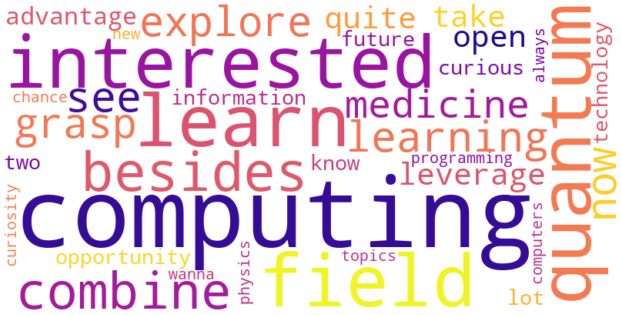}}\hfill
  \subfloat[T7]{\includegraphics[width=0.22\textwidth]{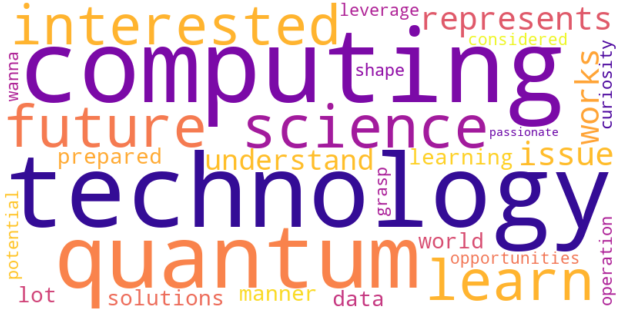}}
  \caption{Per-topic word clouds (English translations). Word size reflects prominence within each topic.}
  \label{fig:wordclouds}
\end{figure*}

To test whether the structure is an artifact of token overlap, we clustered EmbeddingGemma-300M sentence representations (UMAP cosine projection + HDBSCAN). The procedure discovered \textbf{7} clusters with \textbf{11.2\%} noise. Agreement with LDA topics is modest (ARI \(=0.068\), NMI \(=0.163\)), which is expected because clusters capture geometry in a semantic space while LDA partitions vocabulary usage; together they indicate a related but not isomorphic structure. Mean per-cluster silhouette ranges from \(\approx -0.13\) (weak separation) to \(0.50\) (moderate), reflecting heterogeneity typical of short, multi-intent responses. Per-topic word clouds (Fig.~\ref{fig:wordclouds}) provide additional face validity.

\subsection{Takeaway}
Across descriptive and model-based views, \textbf{curiosity-centric language (T1--T2)} aligns with higher observed grades and engagement, whereas \textbf{technology/career framings (T4--T5)} align with lower means. Effects trend as expected but do not reach significance under conservative tests and small \(n\). The embedding-based analysis corroborates a related semantic structure, suggesting the findings are not driven solely by token overlap.

\begin{figure}[t]
  \centering
  \includegraphics[width=\columnwidth]{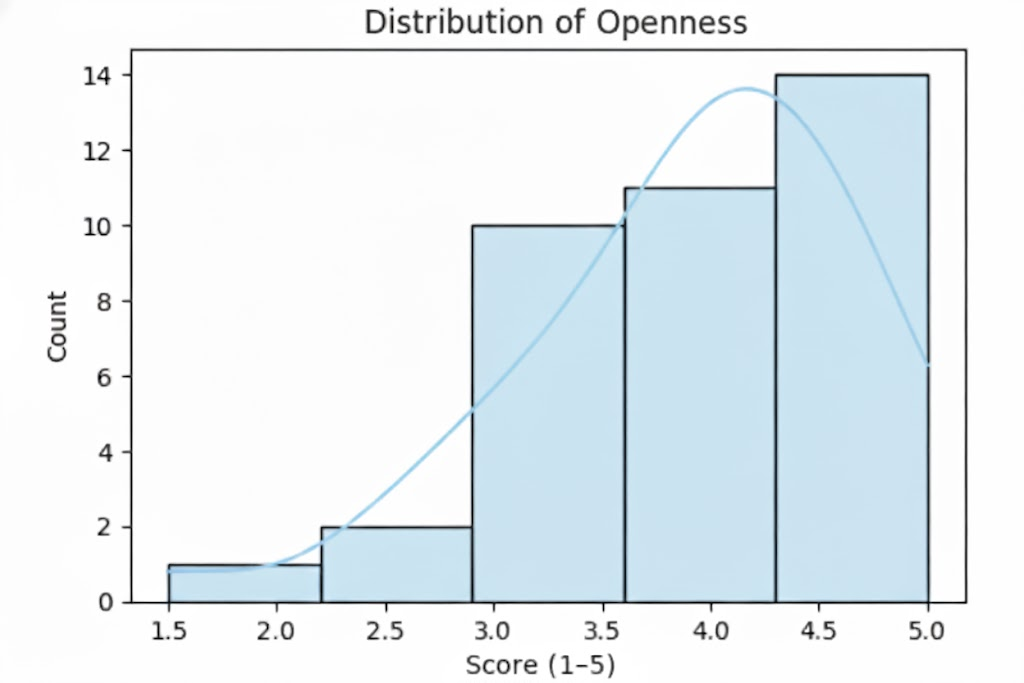}
  \caption{Distribution of BFI-10 Openness (higher = more open). Shown for M2-enrolled students.}
  \label{fig:openness}
\end{figure}

\paragraph*{BFI-10 Openness}
Students in the second module completed BFI-10. The distribution skews high (Fig.~\ref{fig:openness}), consistent with a curiosity-oriented applicant pool. Adding Openness as a covariate in sensitivity analyses did not change the direction of topic effects (small \(\Delta R^2\)); we therefore report it descriptively due to limited coverage.

\section{Discussion}
Short admission responses can be organized into latent themes that map onto intrinsic vs.\ instrumental orientations~\cite{ryan2000sdt,eccles2002expectancy}. In this selective quantum pathway, curiosity-framed language tracks higher grades and attendance than technology/career framings in descriptive analyses. Practically, these signals could guide early support (e.g., mentorship matching) without extra student burden.

Methodologically, combining transparent topics (LDA) with dense embeddings from a small multilingual LM provides converging evidence: LDA aids interpretability; SLM embeddings check stability and capture cross-variety semantic similarity. To our knowledge, this is one of the first studies applying both topic modeling and small language model embeddings to Spanish-language motivation responses in STEM pipelines, addressing an underserved linguistic context.

At the same time, the small outcome sample and heterogeneous grading formulas mean that our findings should be interpreted as exploratory rather than confirmatory. The study does not aim to establish predictive accuracy for high-stakes decisions, but to examine whether meaningful latent structures in motivation language can be detected at all in a realistic, resource-constrained setting. Even under these constraints, the consistency of descriptive trends and the semantic coherence of both LDA topics and SLM clusters provide substantive value for designing future, better-powered studies.

\section{Limitations and Future Work}
This study has several limitations. First, small outcome samples (M1 \(n=23\); M2 \(n=36\)) limit statistical power, so non-significant results cannot be taken as evidence of null effects. Second, cohort-specific grading formulas introduce heterogeneity that complicates cross-cohort comparisons. Third, LDA’s bag-of-words representation ignores syntax and finer-grained discourse markers, and brief Spanish responses (ES-PE/ES-LA) may bias topic discovery.

Future work includes: (i) larger intakes with standardized grading within cohorts; (ii) pre-registered predictive evaluations that explicitly quantify out-of-sample performance; and (iii) fairness analyses across demographic subgroups to ensure that text-based signals do not amplify existing inequities. As small language models continue to improve, lightweight multilingual encoders may enable richer representations of student discourse in under-resourced educational contexts.

\section{Conclusion}
This exploratory study shows that short Spanish motivation statements from applicants to a selective quantum computing pathway can be organized into latent themes that align with intrinsic vs.\ instrumental orientations. Descriptively, curiosity- and learning-oriented language is associated with higher grades and attendance than technology- and career-oriented framings, though our outcome samples are small and the resulting statistical evidence is limited, calling for replication with larger cohorts.

Our main contribution is to demonstrate a portable hybrid pipeline---combining LDA with small multilingual language model embeddings---that can be applied in resource-constrained educational settings to better understand student motivation. The results suggest that brief admission responses are a promising, yet currently under-used, source of information for early support in rigorous STEM pipelines, particularly in Spanish-speaking contexts. Confirming and extending these patterns will require larger samples and pre-registered predictive studies, but the present findings provide an initial, data-driven motivation for such work.

\section*{Acknowledgment}
We thank \emph{QuantumHub Perú} for offering free, selective early-pipeline programs in quantum computing that broaden access to high-level STEM learning. We are grateful to the instructional staff, mentors, and volunteers who designed and delivered Modules~1--2. Above all, we thank participating students and their parents/guardians for their trust and for providing informed consent for de-identified data use. Opinions are those of the authors and do not necessarily reflect the views of QuantumHub Perú.

\appendices
\section{Embedding-based Clustering Details}
\begin{figure}[t]
  \centering
  \includegraphics[width=\columnwidth]{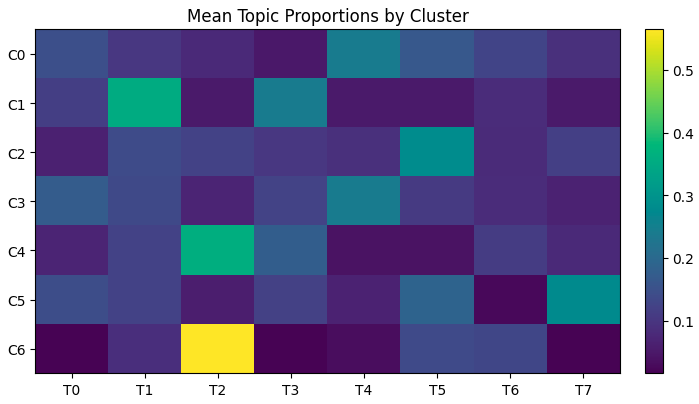}
  \caption{Cluster--topic alignment: mean LDA proportions per embedding-based cluster.}
  \label{fig:heatmap-clu-topic}
\end{figure}

\begin{figure}[t]
  \centering
  \includegraphics[width=\columnwidth]{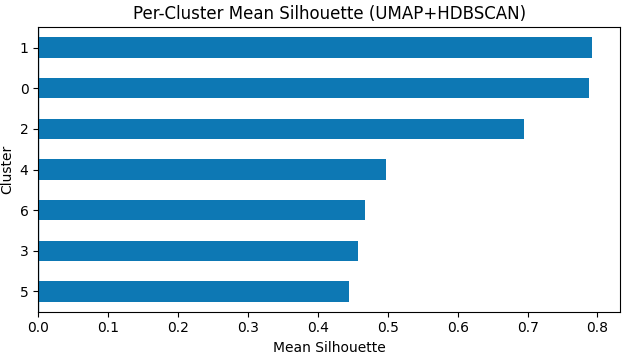}
  \caption{Per-cluster mean silhouette in the UMAP space (higher is better).}
  \label{fig:silhouette-per-cluster}
\end{figure}

We embedded each response with \emph{EmbeddingGemma-300M}~\cite{embeddinggemma-docs,gemma2}, projected with UMAP (cosine; \(n_{\text{neighbors}}=5\), \(n_{\text{components}}=30\))~\cite{mcinnes2018umap}, and clustered via HDBSCAN (EOM; \(\text{min\_cluster\_size}=8\), \(\text{min\_samples}=1\))~\cite{campello2015hdbscan}. The pipeline produced \textbf{7} clusters with \textbf{11.2\%} noise. Figure~\ref{fig:heatmap-clu-topic} shows mean LDA topic proportions per cluster (alignment heatmap); Figure~\ref{fig:silhouette-per-cluster} shows per-cluster mean silhouette.

\begin{thebibliography}{00}

\bibitem{ryan2000sdt}
R.~M. Ryan and E.~L. Deci, ``Self-determination theory and the facilitation of intrinsic motivation, social development, and well-being,'' \emph{American Psychologist}, vol.~55, no.~1, pp.~68--78, 2000.

\bibitem{eccles2002expectancy}
J.~S. Eccles and A.~Wigfield, ``Motivational beliefs, values, and goals,'' \emph{Annual Review of Psychology}, vol.~53, pp.~109--132, 2002.

\bibitem{blei2003lda}
D.~M. Blei, A.~Y. Ng, and M.~I. Jordan, ``Latent Dirichlet Allocation,'' \emph{Journal of Machine Learning Research}, vol.~3, pp.~993--1022, 2003.

\bibitem{mcinnes2018umap}
L.~McInnes, J.~Healy, N.~Saul, and L.~Großberger, ``UMAP: Uniform Manifold Approximation and Projection,'' \emph{Journal of Open Source Software}, vol.~3, no.~29, p.~861, 2018.

\bibitem{campello2015hdbscan}
R.~J.~G.~B. Campello, D.~Moulavi, and J.~Sander, ``Hierarchical density estimates for data clustering, visualization, and outlier detection,'' \emph{ACM Transactions on Knowledge Discovery from Data}, vol.~10, no.~1, pp.~1--51, 2015.

\bibitem{embeddinggemma-docs}
Google, ``EmbeddingGemma,'' developer documentation, 2025. [Online]. Available: \url{https://ai.google.dev/gemma/docs/embeddinggemma}

\bibitem{gemma2}
Gemma Team, ``Gemma~2: Improving Open Language Models at a Practical Scale,'' \emph{arXiv:2408.00118}, 2024.


\bibitem{gragera2025global}
M.~Gragera-Garcés, A.~Hernández, and J.~Molina, ``Introducing quantum computing to high-school curricula: A global perspective,'' \emph{arXiv:2505.14809}, 2025.

\bibitem{maldonadoromo2023quantum}
A.~Maldonado-Romo and L.~Yeh, ``Quantum computing online workshops and hackathon for Spanish speakers: A case study,'' \emph{arXiv:2302.12119}, 2023. (Also in: \emph{Proc. IEEE QCE 2022}, pp.~709--717). Available: \url{https://arxiv.org/abs/2302.12119}

\bibitem{gemma3report}
Gemma Team, ``Gemma 3 Technical Report,'' \emph{arXiv:2503.19786}, 2025. Available: \url{https://arxiv.org/abs/2503.19786}

\bibitem{embeddinggemma2025}
H.~S. Vera \emph{et al.}, ``EmbeddingGemma: Powerful and Lightweight Text Representations,'' \emph{arXiv:2509.20354}, 2025. Available: \url{https://arxiv.org/abs/2509.20354}

\end{thebibliography}
\end{document}